\documentclass{article}
\pdfoutput=1

\usepackage{arxiv}

\usepackage[utf8]{inputenc} % allow utf-8 input
\usepackage[T1]{fontenc}    % use 8-bit T1 fonts
\usepackage{hyperref}       % hyperlinks
\usepackage{url}            % simple URL typesetting
\usepackage{amsfonts}       % blackboard math symbols
\usepackage{microtype}      % microtypography
\usepackage{graphicx}
\usepackage{mathtools}
\usepackage{gensymb}
\title{In-situ AC-hysteresis measurements of SPD-processed Cu$_{20}$(Fe$_{15}$Co$_{85}$)$_{80}$}

\author{
  Martin St\"uckler$^1$, Stefan Wurster, Reinhard Pippan, Andrea Bachmaier \\
  Erich Schmid Institute of Materials Science, Austrian Academy of Sciences\\
  Jahnstra{\ss}e 12, 8700 Leoben, Austria \\
  $^1$\texttt{martin.stueckler@oeaw.ac.at} 
}

\begin{document}
\maketitle

\begin{abstract}
The changes of magnetic properties upon heat treatment of a metastable supersaturated solid solution processed by severe plastic deformation are investigated by in-situ AC-hysteresis measurements. Data are analyzed in the framework of dynamic loss theory, with correlative investigations of the microstructural properties. The evolution of hysteresis upon annealing points out that the single-phase supersaturated solid solution remains stable up to 400\degree C, then hindering of domain wall motion sets in at this temperature. At 600\degree C, a multi phase microstructure is present, causing a significant increase in coercivity.
\end{abstract}

% keywords can be removed

	\section{Introduction}
	With high-pressure torsion (HPT), a technique of severe plastic deformation (SPD), it is possible to form bulk supersaturated solid solutions with grain sizes in the nanocrystalline regime \cite{valiev2000bulk, kormout2017deformation}. It has been shown, that with different ratios of Co to Cu, huge reductions in the coercivity can be achieved by means of SPD, resulting in soft magnetic materials \cite{stuckler2019magnetic, stuckler2020mfm}. In the present study, Fe is added to lower the magnetocrystalline anisotropy to further reduce the coercivity \cite{kuhrt1992formation}. To study the evolution of magnetic properties as a function of temperature, the hysteresis is recorded during in-situ annealing treatments with a concomitant investigation of the dynamic magnetic behavior for certain temperatures. The resulting data are discussed in the framework of dynamic loss theory and correlated to the evolving microstructure.
	\section{Experimental}
	Powders (Fe: MaTeck 99.9\% -100+200 mesh; Co: GoodFellow 99.9\% 50-150 $\mu$m; Cu: Alfa Aesar -170+400 mesh 99.9\%) were mixed and hydrostatically consolidated in Ar-atmosphere. A coin-shaped specimen (diameter: 8~mm; thickness: 1~mm) was processed by two subsequent steps of HPT deformation (100 turns at 300\degree C; 50 turns at room temperature (RT)), as described elsewhere\cite{stuckler2020mfm}. The sample was further processed into a ring shaped specimen and equipped with 68 primary windings and 61 secondary windings (Cu-wire; diameter: 0.315~mm and 0.200~mm, respectively). Electrical isolation between the sample and the windings was maintained with a high temperature adhesive (Minco FortaFix Autostic FC8). To apply the magnetic field, a sinusoidal current (4~A; 5-1000~Hz) was applied to the primary winding with a KEPCO BOP 100-4M power supply, according to [eq.~\ref{eq:H}]. The voltage induced in the secondary windings [eq.~\ref{eq:B}] was measured with a National Instruments BNC-2110 terminal block. Data processing was carried out with LabView (version 14.0.1f3). The hysteresis measurement is described in more detail in Ref.~\cite{turtelli2016hysteresis}. For in-situ measurements, the sample was clamped between two Cu-blocks and heated by cartridge heaters (hotset hotrod HHP HT4030504). The temperature was measured by a K-type thermocouple, close to the samples' position to ensure the hysteresis measurements take place at $\pm$5\degree C of the target temperature. To maintain a homogeneously heated sample, measurements were started 15~min after stabilization of the target temperature. In-situ measurements were performed in a customized vacuum chamber to prevent oxidation, maintaining a pressure below 10$^{-2}$~mbar during the whole experiment.\newline
	Microstructural investigations were performed by X-ray diffraction using Co-K$_\alpha$ radiation (XRD; Bruker D2-Phaser) and scanning electron microscopy (SEM; Zeiss LEO1525) in backscattered electron (BSE) mode. The composition was determined by an energy dispersive X-ray spectroscopy (EDS; Bruker XFlash 6\textbar 60) system.
	\section{Results and discussion} 
\begin{figure}
\begin{center}
\includegraphics[width=0.5\linewidth]{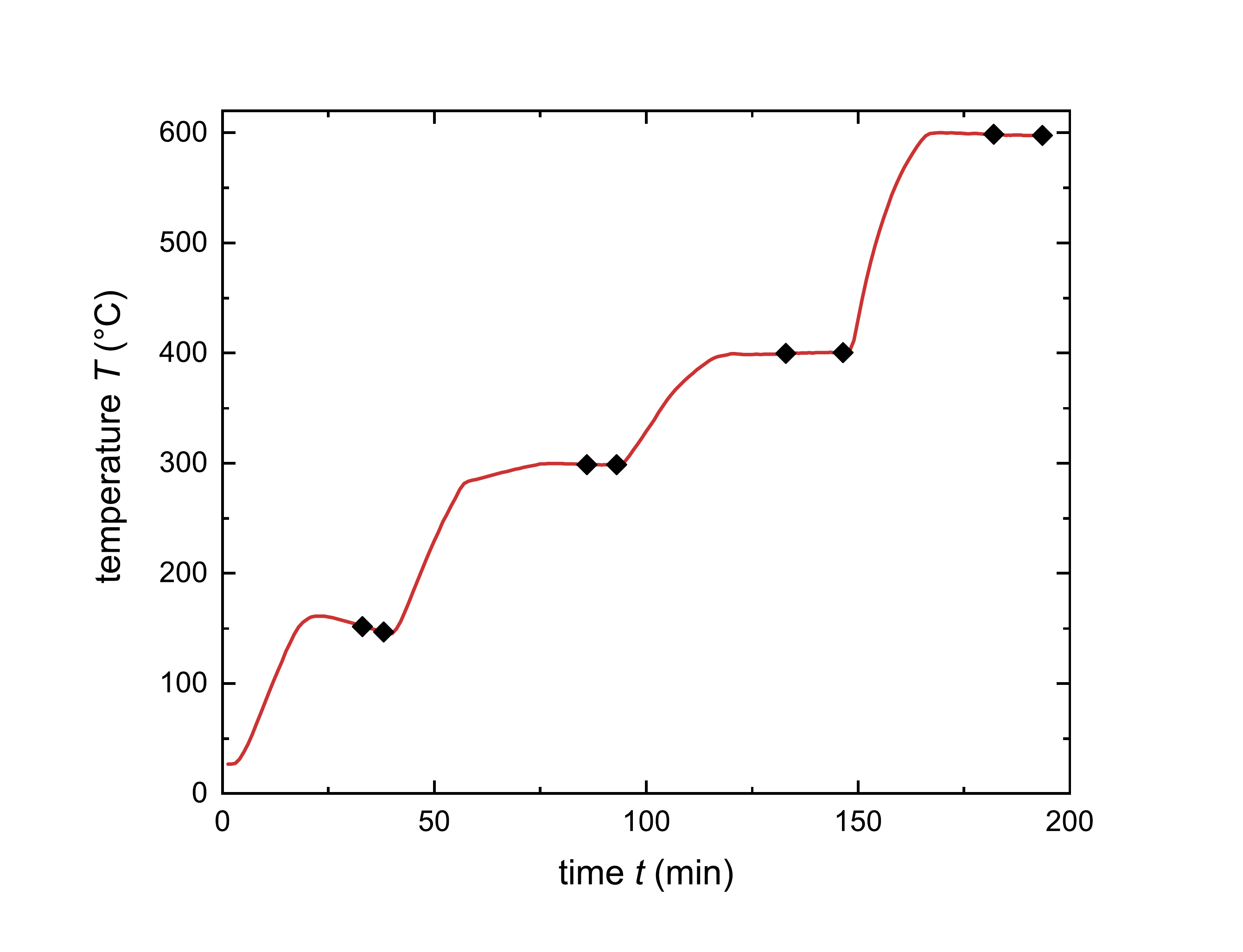}
\end{center}
\caption{Temperature $T$ during the in-situ experiment as a function of time $t$. Black diamonds mark the periods, in which hysteresis measurements are performed.}
\label{fig:T_t}
\end{figure}

\begin{figure}
		\begin{center}
		\includegraphics[width=0.5\linewidth]{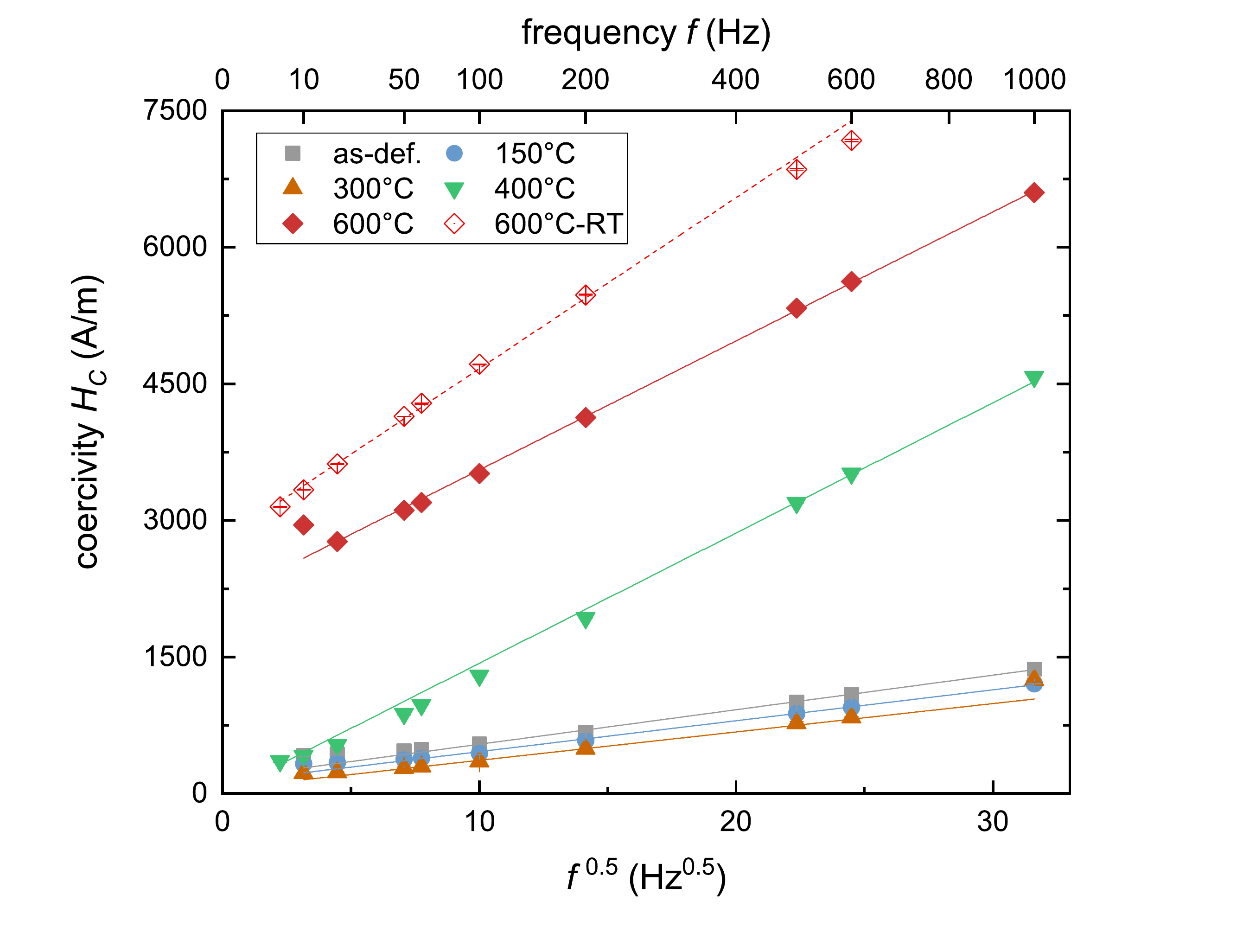}
		\end{center}
		\caption{Coercivity plotted versus the square-root of frequency as recorded during temperature treatment. The lines represent the fits to [eq.~\ref{eq:HC_ac}], taking only anomalous dynamic losses into account.}		
		\label{fig:coerc}
		\end{figure}

\begin{figure}
		\begin{center}
		\includegraphics[width=0.5\linewidth]{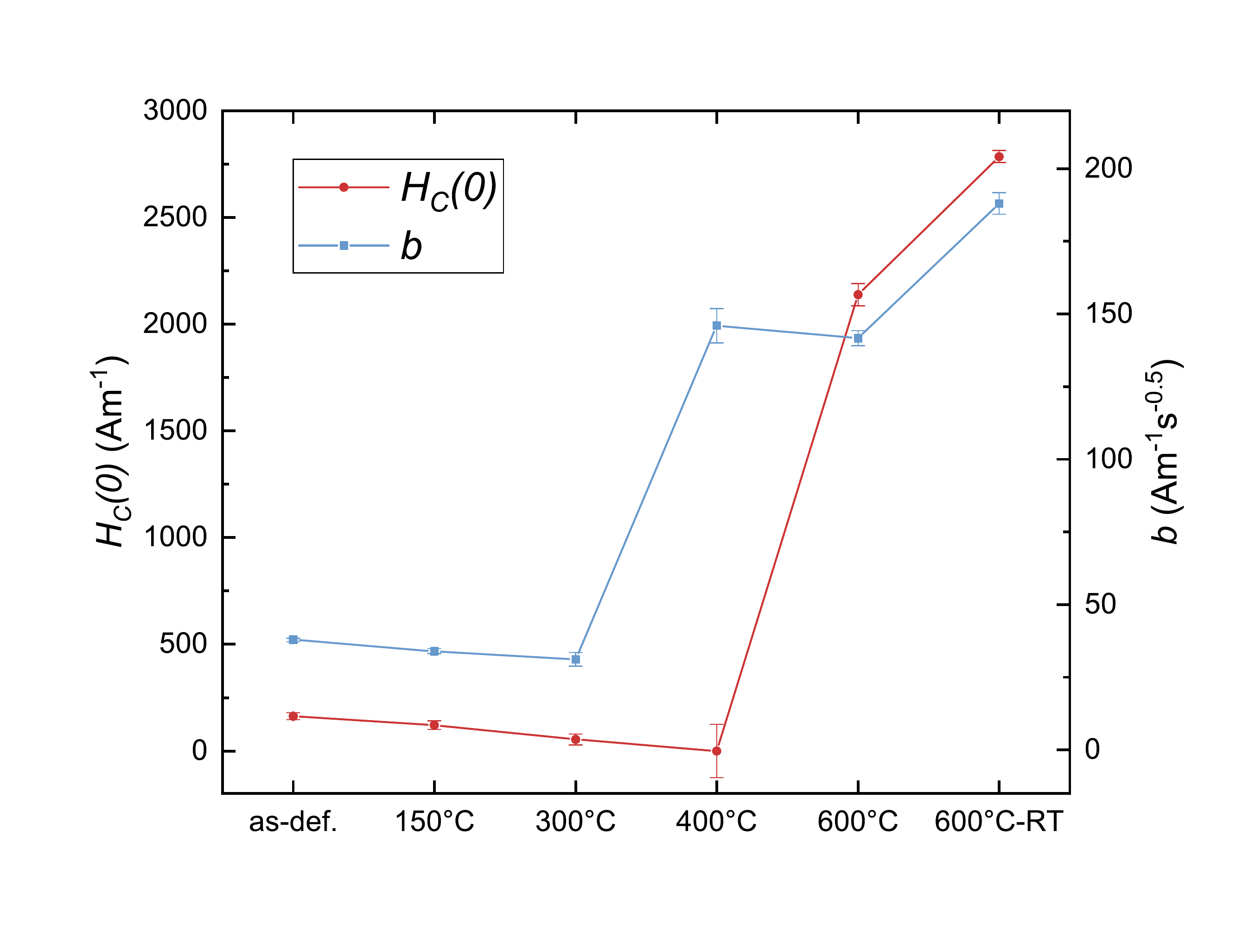}
		\end{center}
		\caption{Static coercivity $H_C(0)$ and anomalous loss parameter $b$, determined according to [eq.~\ref{eq:HC_ac}].}
		\label{fig:fit}
		\end{figure}

\begin{figure}
		\begin{center}
		 \begin{tabular}{cc}
		\includegraphics[width=0.3\linewidth]{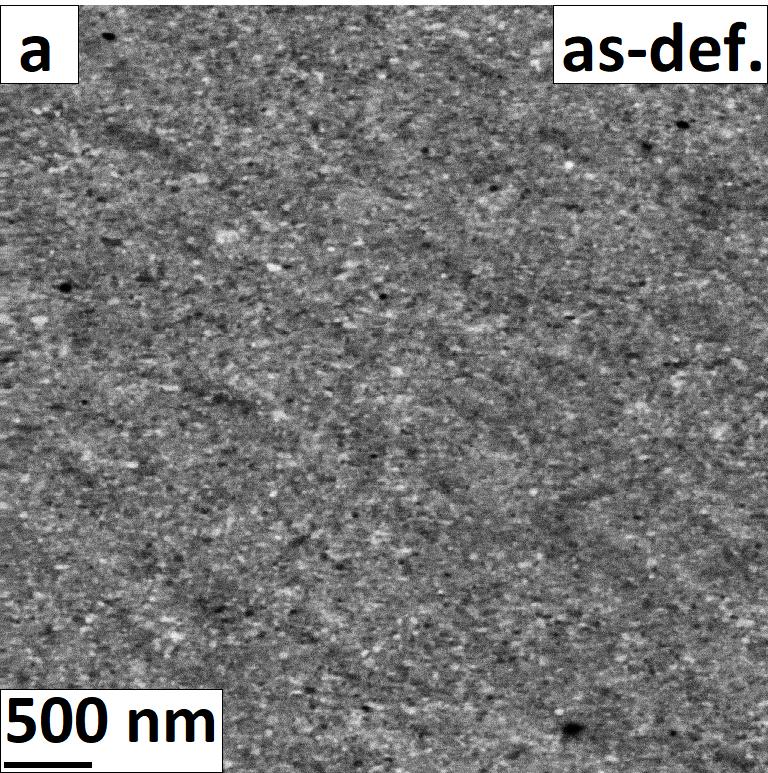} &
		\includegraphics[width=0.3\linewidth]{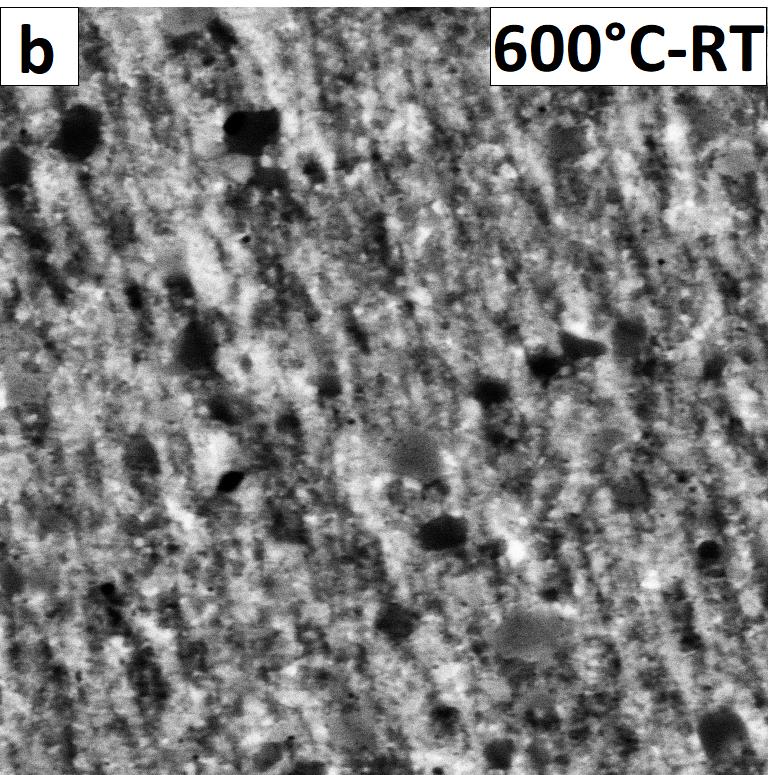}
		\end{tabular}		
		\end{center}
		\caption{SEM-BSE images of a second specimen in as-deformed state (a) and the sample after in-situ heat-treatment (600\degree C-RT; (b)) in tangential sample direction. The scale bar in (a) applies to both images.}
		\label{fig:SEM}
		\end{figure}

\begin{figure}
		\begin{center}
		\includegraphics[width=0.5\linewidth]{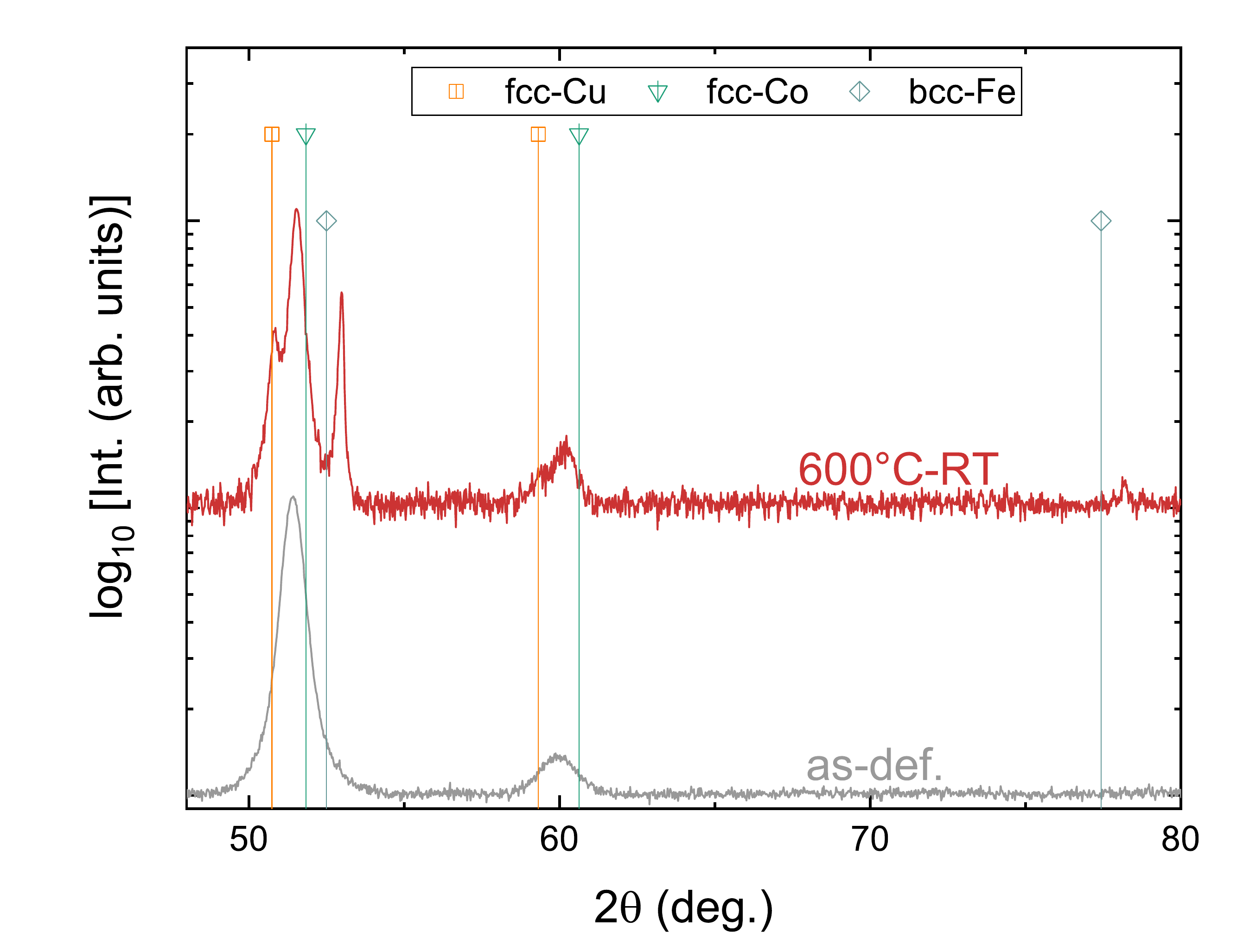}
		\end{center}
		\caption{XRD pattern of the sample after in-situ heat treatment (600\degree C-RT) in comparison with a second specimen in as-deformed state, measured with Co-K$_\alpha$ radiation.}
		\label{fig:xrd}
		\end{figure}

	 The magnetic field $H$ is controlled by the number of primary windings $N_p$, the applied current $I$ and the mean diameter $d_m$ [eq.~\ref{eq:H}], resulting in the present case in a maximum magnetic field of 11.9~kAm$^{-1}$ ($d_{outer}$=8.76~mm; $d_{inner}$=5.76~mm).
	\begin{subequations}
	\begin{equation}
	\label{eq:H}
	H=\frac{N_p \cdot I}{\pi \cdot d_m}
	\end{equation}
	\begin{equation}
	d_m=\frac{d_{outer}+d_{inner}}{2}
	\end{equation}
	\end{subequations}
	Eq.~\ref{eq:B} gives the magnetic induction $B$ as a function of the time dependent induced voltage $U_{ind}(t)$, the number of secondary windings $N_s$ and the cross-sectional area $A$ of the ring-core (here: $A$=0.707~mm$^2$).
	\begin{equation}
	\label{eq:B}
	B=\frac{\int U_ {ind}(t) dt}{N_s \cdot A}
	\end{equation}
	\begin{figure*}	
		\begin{center}
			\begin{tabular}{ccc}
	\includegraphics[width=0.33\linewidth]{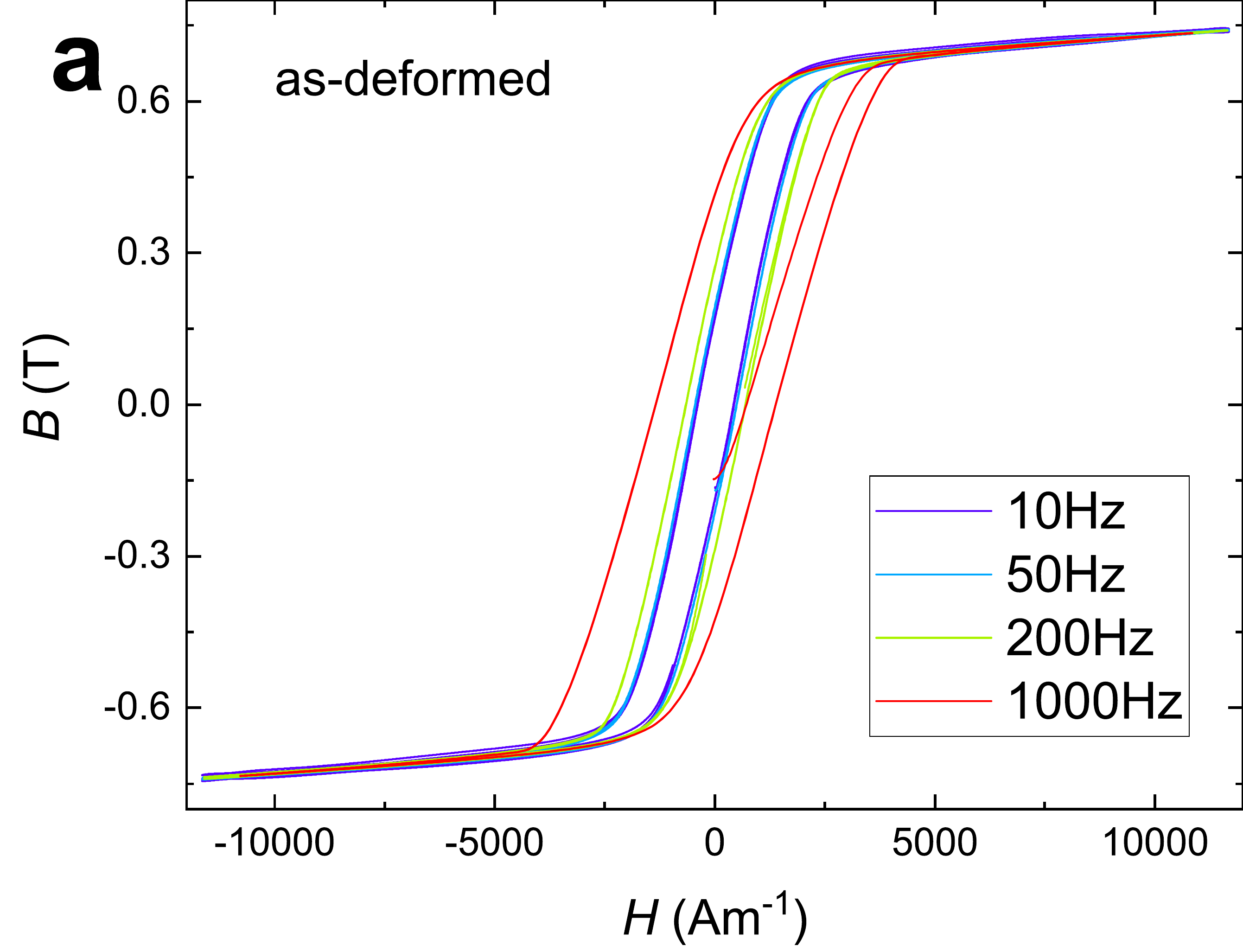}
	&
	\includegraphics[width=0.33\linewidth]{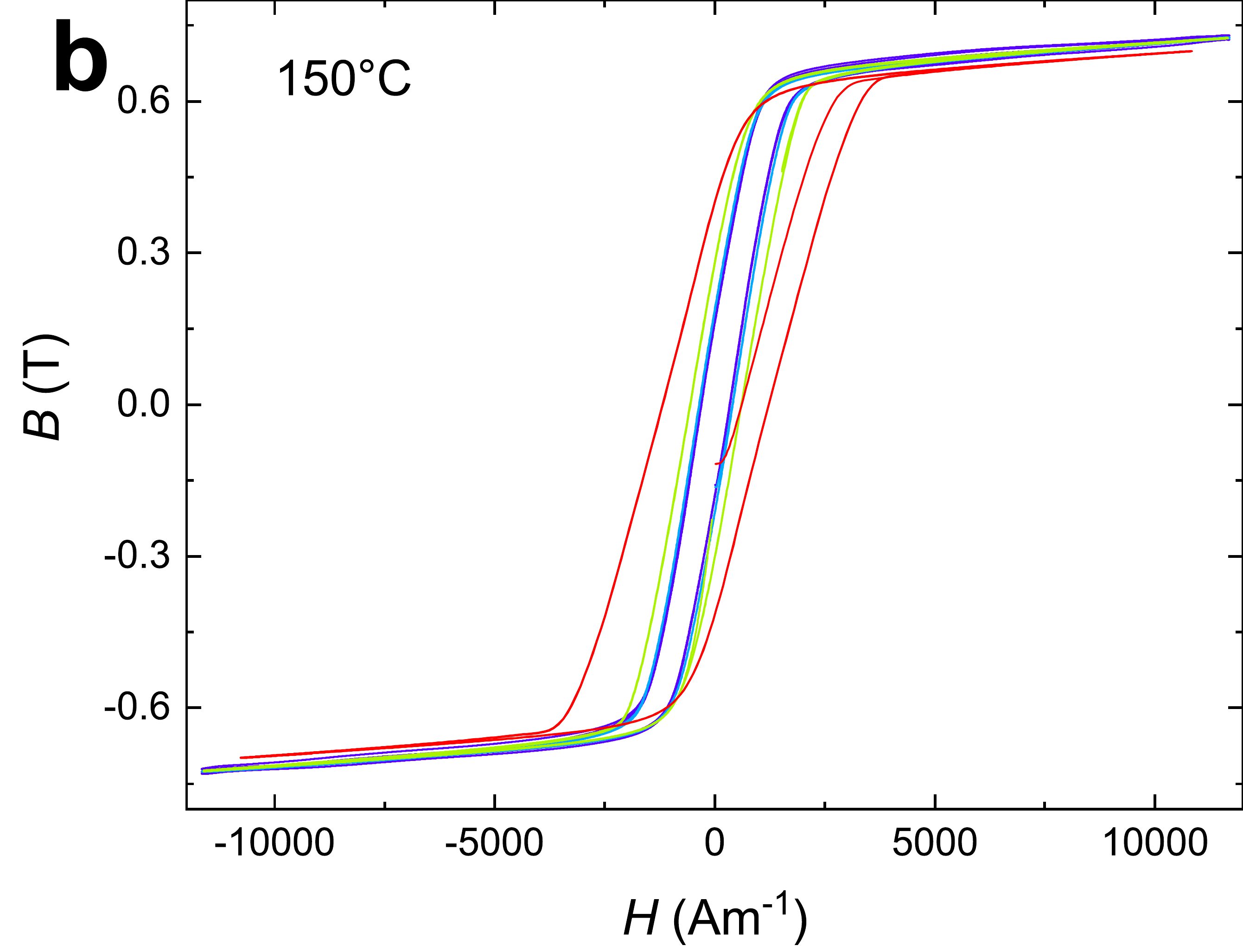}
	&
	\includegraphics[width=0.33\linewidth]{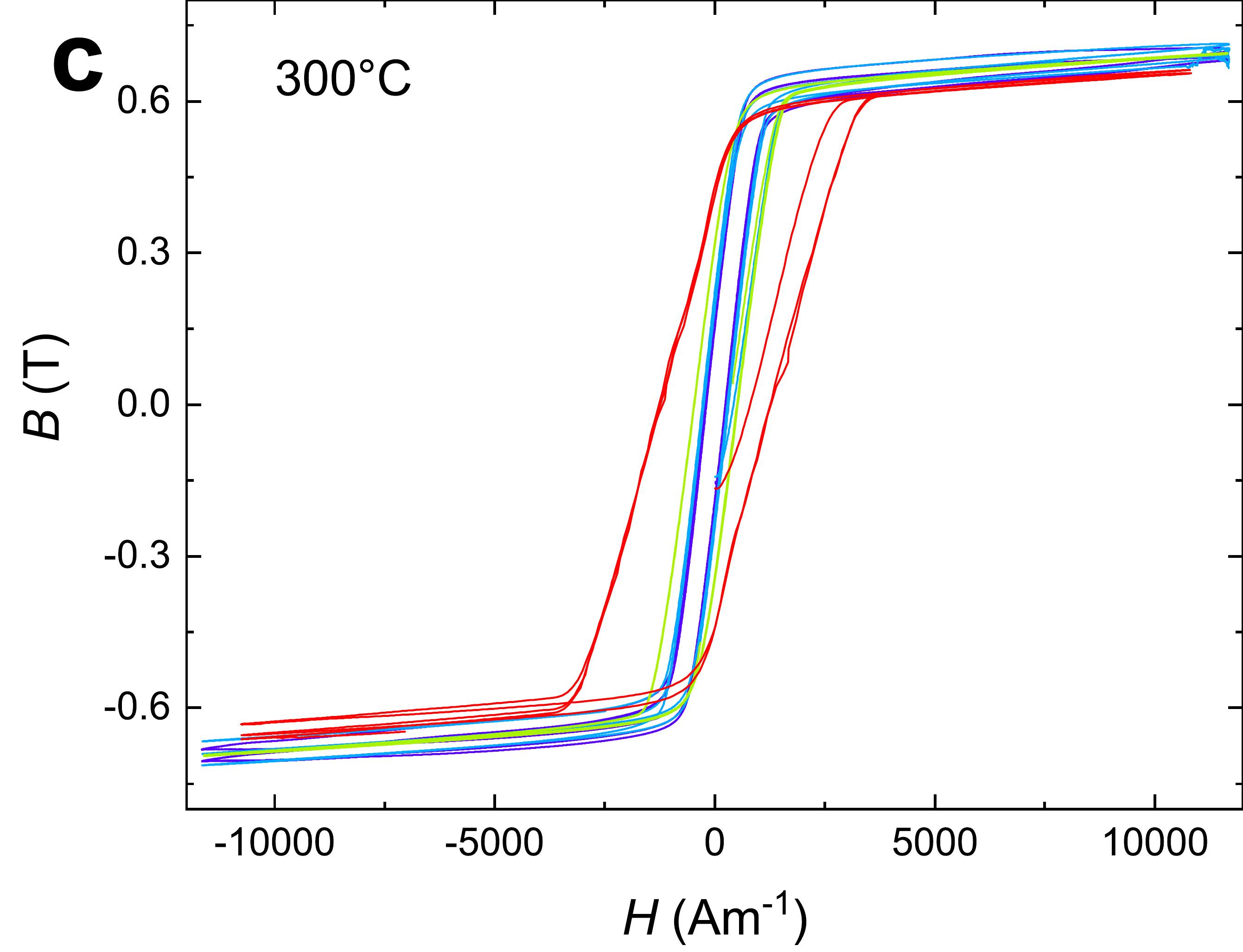}
	\\
	\includegraphics[width=0.33\linewidth]{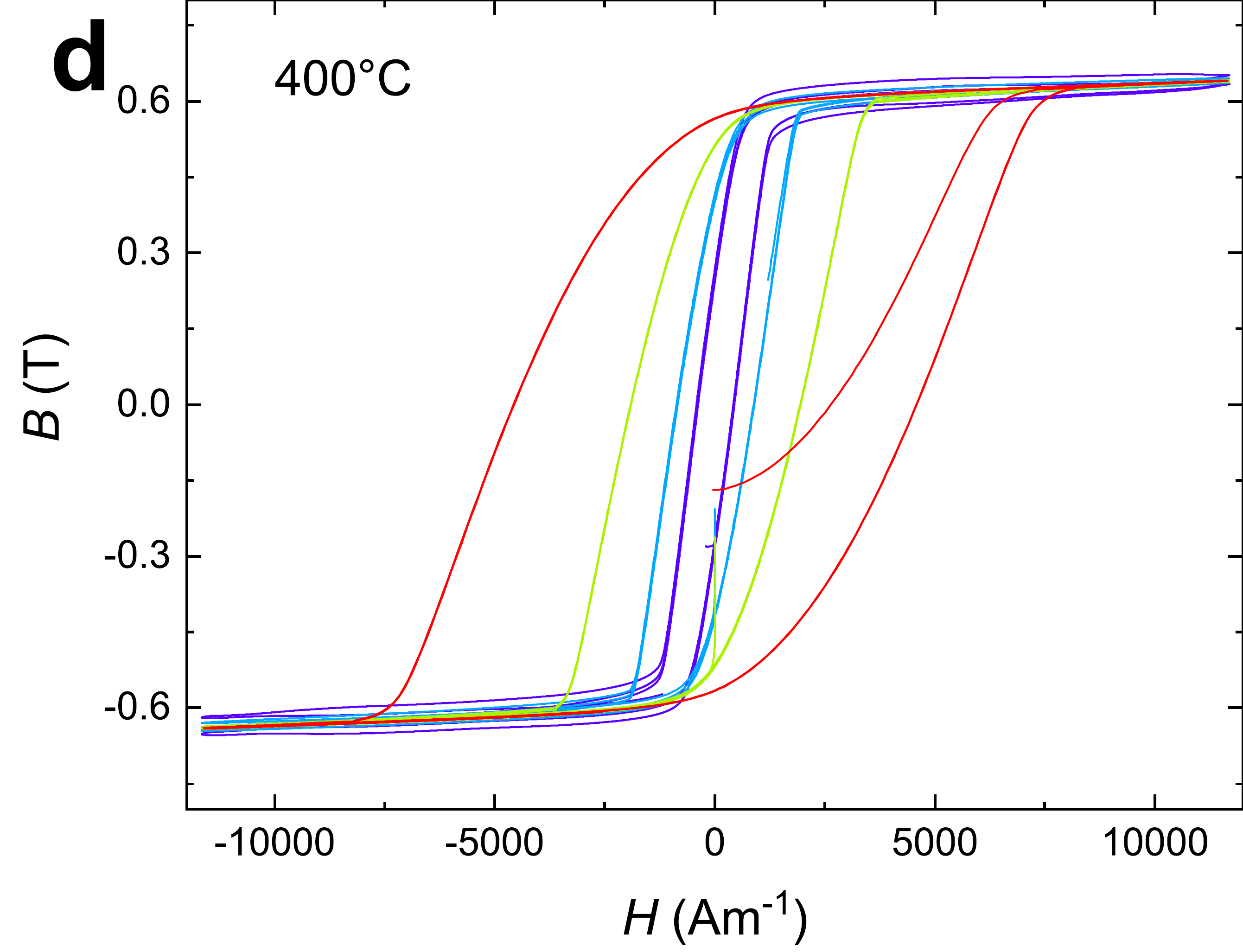}
	&
	\includegraphics[width=0.33\linewidth]{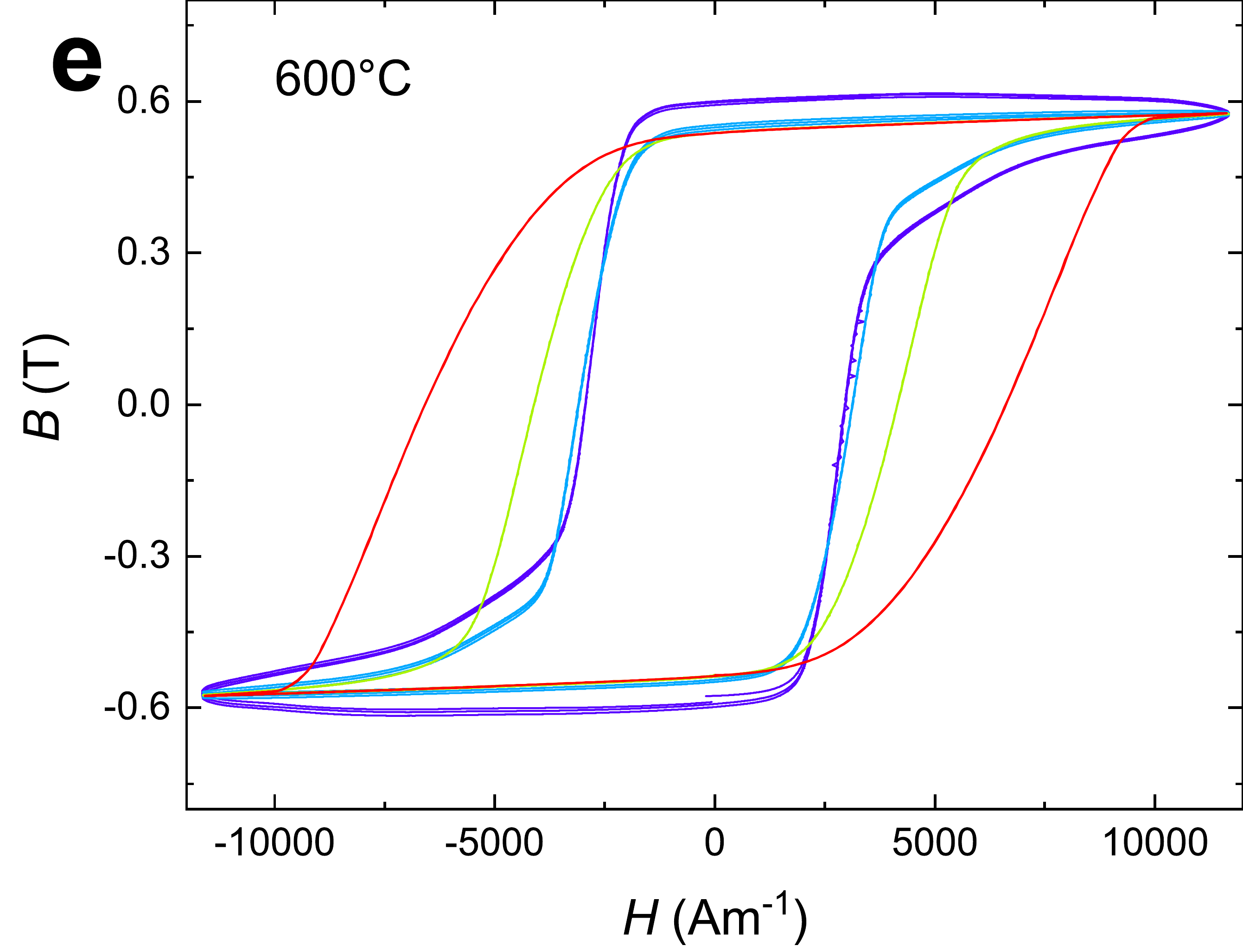}
	&
	\includegraphics[width=0.33\linewidth]{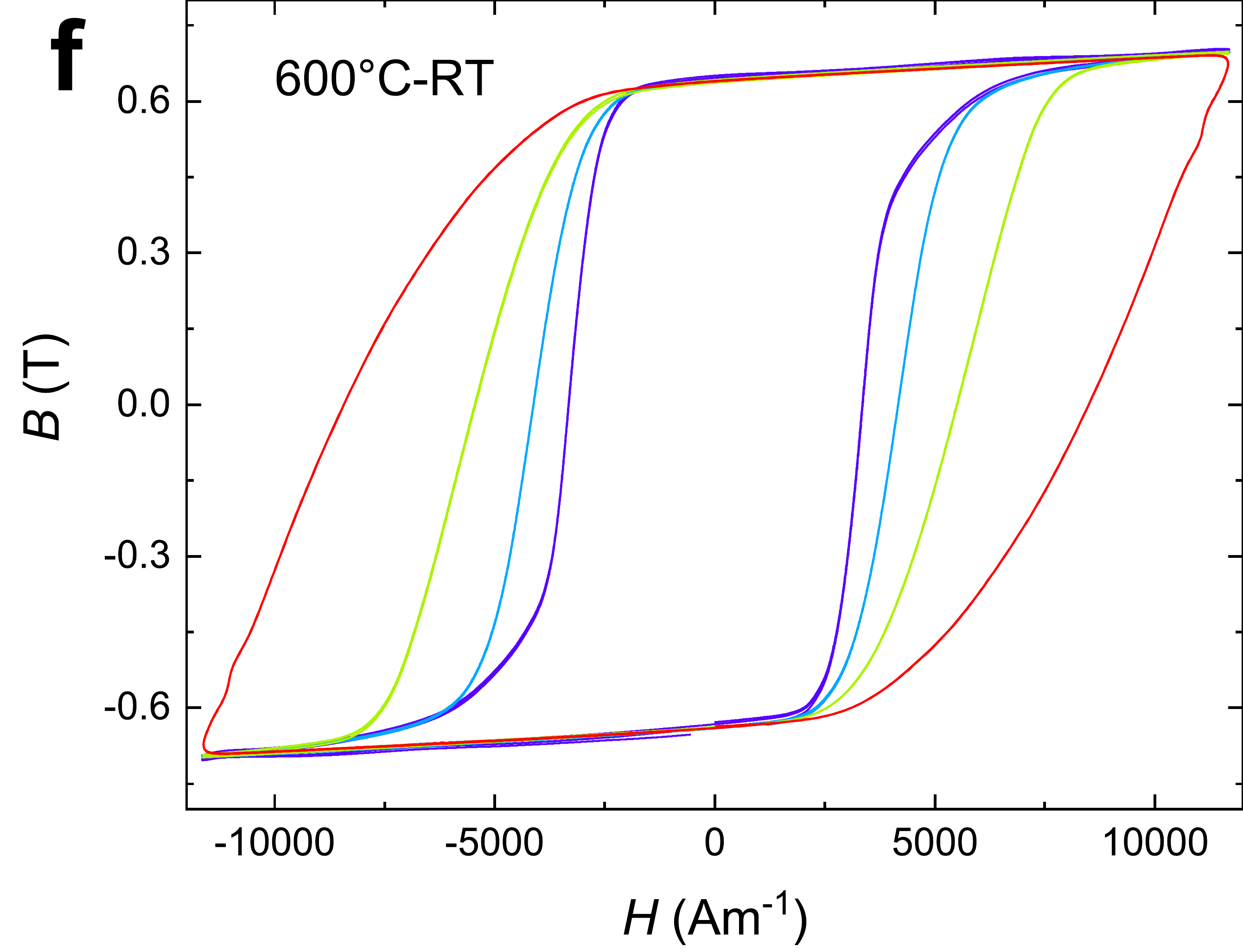}
			\end{tabular}
	\end{center}
	\caption{Evolution of in-situ AC-hysteresis curves according to the temperature treatment in fig.~\ref{fig:T_t}. The as-deformed state is measured at RT (a). After in-situ temperature treatment at 150\degree C (b), 300\degree C (c), 400\degree C (d) and 600\degree C (e), the specimen is measured again after cooling down to room temperature (600\degree C-RT; (f)). The legend in (a) applies to all diagrams.}\label{fig:hyst}
	\end{figure*}
		The specimen is measured in the as-deformed state at RT, shown in fig.~\ref{fig:hyst}(a). It can be seen, that saturation is achieved and the area of the hysteresis loop rises slightly with increasing frequency, indicating low eddy-current losses. 
		The in-situ temperature treatment was performed according to the thermal profile shown in fig.~\ref{fig:T_t}. Measurements were started after settling the target temperature for 15~min, since similarly processed Co-Cu samples showed the majority of microstructural changes happening only during a short time period after reaching the target temperature \cite{wurster2020GMR}. The time stamps of the measurements are represented by the black diamonds in fig.~\ref{fig:T_t}. Hysteresis are measured at 150\degree C (fig.~\ref{fig:hyst}(b)), 300\degree C (fig.~\ref{fig:hyst}(c)), 400\degree C (fig.~\ref{fig:hyst}(d)) and 600\degree C (fig.~\ref{fig:hyst}(e)). For the 150\degree C measurement, the temperature exceeded 160\degree C for a short period of time, but it was shown that the microstructure does not change in this temperature range since only relief of internal stresses takes place to a small extend \cite{stueckler2020jac}. The hysteresis exhibits similar shapes up to 300\degree C, but at 400\degree C the frequency behavior changes significantly, which is clearly visible in the 1000~Hz measurement. The area of the 1000~Hz hysteresis increases again in the 600\degree C-state. The measurement of the in-situ treated specimen is repeated after cooling down to RT (fig.~\ref{fig:hyst}(f); 600\degree C-RT). For the 600\degree C-RT state saturation is not achieved in the 1000~Hz measurement. This means, the measured hysteresis represents a minor loop and the coercivity cannot be evaluated for this frequency. Lower frequencies reveal a larger area of the hysteresis loop with respect to the 600\degree C-state, arising most likely from the temperature dependence of the magnetocrystalline anisotropy according to Brukhatov-Kirensky \cite{brukhatov1937}.\newline
		In the following, a quantitative analysis on the evolution of the coercivity is carried out. Fig.~\ref{fig:coerc} shows the measured coercivities as a function of frequency. To disentangle the (static) intrinsic magnetic properties from the dynamic losses, [eq.~\ref{eq:HC_ac}] is fitted to the data\cite{bertotti1988prediction, fiorillo2004measurement, hilzinger2013magnetic}.
		\begin{equation}
		H_C(f)=H_C(0)+ b \cdot \sqrt{f} + c \cdot f
		\label{eq:HC_ac}
		\end{equation}
		In [eq.~\ref{eq:HC_ac}], the dynamic loss is separated into anomalous-loss $b$, caused by domain wall motion, and eddy current loss $c$, which is mainly controlled by the conductivity\cite{fiorillo2016}. Since the conductivity of SPD-processed materials is significantly lowered with respect to coarse-grained materials \cite{cubero2015high}, we assume the dynamic losses being mainly controlled by anomalous losses and therefore neglect the third term in [eq.~\ref{eq:HC_ac}]. In fig.~\ref{fig:coerc}, the measured coercivities are plotted versus the square-root of frequency, showing a linear scaling with $\sqrt{f}$, confirming the aforementioned statement. The results from linearly fitting the data are shown in fig.~\ref{fig:fit}. Diminishing static coercivity, as well as anomalous loss, can be identified between the as-deformed, 150\degree C-  and 300\degree C-annealed state. For SPD-processed Cu-Co and Cu-Fe-Co, a diminishing defect density was reported in this temperature window, but no apparent changes in the grain size have been determined \cite{stuckler2020mfm, stueckler2020jac}. The decreasing coercivity between RT and 400\degree C might therefore originate from a decrease in the magnetoelastic anisotropy constant\cite{shen2005soft} $K_{el} \propto \sigma \cdot \lambda$, with the residual stress $\sigma$ and the magnetostriction $\lambda$. It should be stated, that the coercivity is further lowered by the temperature dependence of the magnetocrystalline anisotropy, as already mentioned \cite{brukhatov1937}. A huge jump in $H_C(0)$ can be noticed at 600\degree C, indicating a large microstructural variation, such as grain growth. Large microstructural variations have been determined in similar materials at this temperature \cite{stuckler2020mfm, stueckler2020jac}. A further increase in coercivity is visible for the 600\degree C-RT state, which is again traced back to temperature dependence.\newline
				The anomalous loss parameter $b$ is closely related to the microstructure, and takes into account the energy needed for domain wall motion, which can be increased due to pinning at lattice defects or grain / phase boundaries. The increase in $b$ at 400\degree C therefore indicates the formation of pinning centers accelerating domain wall motion \cite{houze1967domain, overshott1976use}, which rushes ahead the demixing of the microstructure at 600\degree C. For the 600\degree C-state, the anomalous loss parameter stays rather constant.\newline
		The microstructure of the in-situ heat-treated sample is investigated by SEM and XRD and compared to the initial (as-deformed) state. For this purpose a second sample is fabricated, representing the as-deformed state. Fig.~\ref{fig:SEM} shows SEM images of both samples. The as-deformed state (fig.~\ref{fig:SEM}(a)) shows a highly homogeneous, nanocrystalline microstructure. In the 600\degree C-RT state (fig.~\ref{fig:SEM}(b)), a significantly larger grain size in the ultra-fine grained regime is visible. Furthermore, phase contrast indicates a chemical inhomogeneity, showing demixing tendencies. Bright areas indicate high Z and therefore Cu-rich regions, whereas dark-grey or black areas point at the presence of low Z Fe-Co-rich alloy intermetallic phases. EDS-measurements at 20 different spots reveal the composition of the in-situ treated sample to Cu$_{23}$(Fe$_{12}$Co$_{88}$)$_{77}$ (wt.\%; $\pm$1~wt.\%).\newline
		In fig.~\ref{fig:xrd}, the XRD pattern of the in-situ treated sample is shown in comparison to the as-deformed state. In the as-deformed state, only one fcc-pattern is visible, revealing the single-phase crystallographic structure, whereas in the 600\degree C-RT state, three different patterns evolve: the fcc-Cu pattern coincides with the theoretical values, indicating the presence of pure Cu. In contrast, the fcc-Co, as well as the bcc-Fe pattern, show deviations with respect to the theoretical values, suggesting the presence of $\gamma$-(Fe,Co) and $\alpha$-(Fe,Co), in line with the SEM investigations above. %The in-situ treatment therefore represents the evolution from a metastable single-phase supersaturated solid solution towards a demixing of phases according to the thermodynamical equilibrium.
	
	\section{Conclusion}
In-situ AC-hysteresis measurements of SPD-processed Cu$_{20}$(Fe$_{15}$Co$_{85}$)$_{80}$ reveal a persisting soft magnetic behavior up to 400\degree C. The amount of eddy current losses is low by comparison, owing to the high resistivity of SPD-processed materials. At 400\degree C, pinning centers start to form, accelerating domain wall motion and causing an increase in dynamic loss. At 600\degree C, the microstructure has changed from the initial single-phase supersaturated solid solution into a multi phase microstructure according to the thermodynamical equilibrium. The formation of pinning centers rushes ahead this phase change. The results demonstrate the capability of magnetic measurements capturing smallest microstructural changes before they become evident with other techniques.

\section{Acknowledgments}
This project has received funding from the European Research Council (ERC) under the European Union’s Horizon 2020 research and innovation programme (Grant No. 757333). The authors thank R. Neubauer, M. Reiter and M. Kasalo for preparing the sample and the respective holder for in-situ experiment.
\section*{Data Availibility Statement}
The data that support the findings of this study are available from the corresponding author upon reasonable request.
\bibliography{libraryaip}

\begin{thebibliography}{10}

\bibitem{valiev2000bulk}
R.~Z. Valiev, R.~K. Islamgaliev, and I.~V. Alexandrov.
\newblock Bulk nanostructured materials from severe plastic deformation.
\newblock {\em Prog. Mater. Sci.}, 45(2):103--189, 2000.

\bibitem{kormout2017deformation}
K.~S. Kormout, R.~Pippan, and A.~Bachmaier.
\newblock Deformation-induced supersaturation in immiscible material systems
  during high-pressure torsion.
\newblock {\em Adv. Eng. Mater.}, 19(4):1600675, 2017.

\bibitem{stuckler2019magnetic}
M.~St{\"u}ckler, H.~Krenn, R.~Pippan, L.~Weissitsch, S.~Wurster, and
  A.~Bachmaier.
\newblock Magnetic binary supersaturated solid solutions processed by severe
  plastic deformation.
\newblock {\em Nanomaterials}, 9(1):6, 2019.

\bibitem{stuckler2020mfm}
M.~St{\"u}ckler, C.~Teichert, A.~Matkovi{\'c}, H.~Krenn, L.~Weissitsch,
  S.~Wurster, R.~Pippan, and A.~Bachmaier.
\newblock On the magnetic nanostructure of a {Co-Cu} alloy processed by
  high-pressure torsion.
\newblock {\em J. Sci. Adv. Mater. Dev.}, 2020.
\newblock in press.

\bibitem{kuhrt1992formation}
C.~Kuhrt and L.~Schultz.
\newblock Formation and magnetic properties of nanocrystalline mechanically
  alloyed {Fe‐Co}.
\newblock {\em J. Appl. Phys.}, 71(4):1896--1900, 1992.

\bibitem{turtelli2016hysteresis}
R.~S. Turtelli, S.~Hartl, R.~Gr{\"o}ssinger, R.~W{\"o}hrnschimmel,
  D.~Horwatitsch, F.~Spieckermann, G.~Polt, and M.~Zehetbauer.
\newblock Hysteresis and loss measurements on the plastically deformed {Fe}--(3
  wt\%) {Si} under sinusoidal and triangular external field.
\newblock {\em IEEE Trans. Magn.}, 52(5):1--7, 2016.

\bibitem{wurster2020GMR}
S.~Wurster, M.~Stückler, L.~Weissitsch, T.~Müller, and A.~Bachmaier.
\newblock Microstructural changes influencing the magnetoresistive behavior of
  bulk nanocrystalline materials.
\newblock {\em Appl. Sci.}, 10(15):5094, Jul 2020.

\bibitem{stueckler2020jac}
M.~St\"uckler, L.~Weissitsch, S.~Wurster, H.~Krenn, R.~Pippan, and
  A.~Bachmaier.
\newblock Formation of solid solutions of {Cu-Fe-Co} by severe plastic
  deformation.

\bibitem{brukhatov1937}
N.~L. Brukhatov and L.~V. Kirensky.
\newblock The anisotropy of the magnetic energy in single crystals of nickel as
  a function of temperature.
\newblock {\em Phys Z Sowjetunion}, 12(5):601, 1937.

\bibitem{bertotti1988prediction}
G.~Bertotti, F.~Fiorillo, and G.~P. Soardo.
\newblock The prediction of power losses in soft magnetic materials.
\newblock {\em J. Phys. Colloques}, 49(C8):C8--1915, 1988.

\bibitem{fiorillo2004measurement}
F.~Fiorillo.
\newblock {\em Measurement and characterization of magnetic materials}.
\newblock North-Holland, 2004.

\bibitem{hilzinger2013magnetic}
R.~Hilzinger and W.~Rodewald.
\newblock {\em Magnetic materials: fundamentals, products, properties,
  applications}.
\newblock Vacuumschmelze, 2013.

\bibitem{fiorillo2016}
F.~Fiorillo, G.~Bertotti, C.~Appino, and M.~Pasquale.
\newblock {\em Soft Magnetic Materials}, pages 1--42.
\newblock 2016.

\bibitem{cubero2015high}
J.~M. Cubero-Sesin, M.~Arita, and Z.~Horita.
\newblock High strength and electrical conductivity of {Al-Fe} alloys produced
  by synergistic combination of high-pressure torsion and aging.
\newblock {\em Adv. Eng. Mater.}, 17(12):1792--1803, 2015.

\bibitem{shen2005soft}
T.~D. Shen, R.~B. Schwarz, and J.~D. Thompson.
\newblock Soft magnetism in mechanically alloyed nanocrystalline materials.
\newblock {\em Phys. Rev. B}, 72:014431, Jul 2005.

\bibitem{houze1967domain}
G.~L. Houze~Jr.
\newblock Domain-wall motion in grain-oriented silicon steel in cyclic magnetic
  fields.
\newblock {\em J. Appl. Phys.}, 38(3):1089--1096, 1967.

\bibitem{overshott1976use}
K.~Overshott.
\newblock The use of domain observations in understanding and improving the
  magnetic properties of transformer steels.
\newblock {\em IEEE Trans. Magn.}, 12(6):840--845, 1976.

\end{thebibliography}

\end{document}